# CHAPTER 11: Data Analytics for Intermodal Freight Transportation Applications

## ABSTRACT

With the growth of intermodal freight transportation, it is important that transportation planners and decision makers are knowledgeable about freight flow data to make informed decisions. This is particularly true with Intelligent Transportation Systems (ITS) offering new capabilities to intermodal freight transportation. Specifically, ITS enables access to multiple different data sources, but they have different formats, resolution, and time scales. Thus, knowledge of data science is essential to be successful in future ITS-enabled intermodal freight transportation system. This chapter discusses the commonly used descriptive and predictive data analytic techniques in intermodal freight transportation applications. These techniques cover the entire spectrum of univariate, bivariate, and multivariate analyses. In addition to illustrating how to apply these techniques through relatively simple examples, this chapter will also show how to apply them using the statistical software R. Additional exercises are provided for those who wish to apply the described techniques to more complex problems.

## LIST OF AUTHORS

Nathan Huynh
Associate Professor
University of South Carolina
300 Main Street
Columbia, SC 29208
Phone: 803-777-8947
Email: nathan.huynh@sc.edu

Majbah Uddin
Ph.D. Candidate
University of South Carolina
Department of Civil and Environmental Engineering
300 Main Street
Columbia, SC 29208
Telephone: (803) 447-4445
Email: muddin@cec.sc.edu

Chu Cong Minh
Associate Professor
Ho Chi Minh City University of Technology
Department of Bridge and Highway Engineering

268 Ly Thuong Kiet Street, District 10
Ho Chi Minh City, Vietnam
Email: ccminh@hcmut.edu.vn

**SHORT BIOGRAPHIES**

**Nathan Huynh** is an Associate Professor at the University of South Carolina.  He received his Ph.D. and MS in Transportation Engineering from the University of Texas at Austin.  His research interests include intermodal freight terminal design and operations, intermodal freight transportation, and freight transportation planning and logistics.

**Md Majbah Uddin** is a Ph.D. candidate in the Civil and Environmental Engineering Department at the University of South Carolina.  He received a MS in civil engineering with specialization in transportation from the University of South Carolina.  His research interests include transportation planning, freight assignment, and truck safety.

**Chu Cong Minh** is an Associate Professor at the Ho Chi Minh City University of Technology, Vietnam.  He received his Ph.D. in Transportation Engineering from the Nagaoka University of Technology, Japan.  His research interests include terminal design and operations, and transportation planning and management.

# CHAPTER 11

# Data Analytics for Intermodal Freight Transportation Applications

## INTRODUCTION

Intermodal freight transportation is defined as the use of two or more modes to move goods in intermodal containers from an origin to a destination. Intermodal freight transportation consists of three segments, *pre-haul* (transportation of container from shipper to origin terminal), *long-haul* (transportation of container from origin terminal to destination terminal), and *end-haul* (transportation of container from destination terminal to receiver). Typically, the pre-haul and end-haul are carried out by trucks, and the long-haul is carried out via rail, air or water. The transfer of containers from one mode to another is performed at an intermodal terminal.

The beginning of intermodal freight transportation dates back to the 18th century. In the earlier days, when a transfer to another mode is required (e.g., ship to train), the boxes, barrels, or bags in which goods were stored in are unloaded from one mode and then reloaded to another mode using manual labor and primitive equipment. Thus, the transferring process was extremely slow. This process became much more efficient in the mid-1950s with the creation of standardized intermodal containers by Malcom McLean. With the use of containers, the transfer of modes is greatly facilitated by mechanical cranes. Today, at most U.S. ports, quay cranes (also known as ship-to-shore cranes) are capable of unloading or loading up to 40 containers per hour.

The U.S. currently has the largest freight transportation system in the world [1]; it moved, on average, 54 million tons worth nearly $48 billion of freight each day in 2012 and the majority of freight was transported by either truck or rail (67% by truck and 10% by rail). The freight volume is expected to increase to 78 million tons (about 45%) by the year 2040 [2]. In recent years, intermodal transportation is becoming an increasingly attractive alternative to shippers, and this trend is likely to continue as governmental regulatory agencies are considering policies to induce a freight modal shift from road to intermodal to alleviate highway congestion and emissions.

## ITS-Enabled Intermodal Freight Transportation

Intelligent Transportation Systems (ITS) provide new functionalities to intermodal freight transportation that will improve planning, operational efficiency, energy consumption, safety, air emissions, and customer satisfaction. One of the emerging ITS-based systems for intermodal freight transportation is Freight Advanced Traveler Information Systems (FRATIS), which has been implemented in several U.S. cities. Specifically, FRATIS has been applied to use queuing times at ports and real-time traffic information on roadways to optimize truck movements.

**Data Analytics for ITS-Enabled Intermodal Freight Transportation**

With FRATIS and other emerging ITS-based systems providing a multitude of data for intermodal freight transportation, knowledge of data analytics is essential for transportation planners and decision makers in understanding and leveraging this Big Data set. Data analytics is the science of collecting, organizing, and analyzing datasets to identify patterns and draw conclusions, and Big Data refers to the huge amount of available information being generated by multifaceted electronic systems. By taking advantage of the power data analytics, stakeholders can simplify global shipments and arrange for more efficient door-to-door delivery using all types of transport and improve intermodal truck utilization. Data analytics can also be used to visualize the impact of government regulation on the ports and the economy. In summary, data analytics can provide better insight into problems and allow members of the logistics industry to respond more efficiently.

This chapter will discuss the data analytic techniques that are relevant for intermodal freight transportation applications. It will discuss and illustrate the use of descriptive and predictive data analytic techniques. In addition to illustrating how to apply these techniques through relatively simple examples, it will also show how these techniques can be applied using the statistical software R.

**DESCRIPTIVE DATA ANALYTICS**

**Univariate Analysis**

Univariate analysis applies to data sets that consist of a single variable. If the data set consists of continuous variables, then the measures of interest are the central tendency and spread of the variable. These measures can be visualized via a histogram or box plot. Table 11.1 summarizes the measures of interest for continuous variables. For categorical variables, a frequency table can be used to determine either the total count or count percentage of each category. Bar charts can be used to visualize the frequency data.

**Table 11.1** Measures of interest for continuous variables

| Central Tendency | Measure of Dispersion | Visualization Method |
|---|---|---|
| Mean | Range | Histogram |
| Median | Quartile | Box plot |
| Mode | Interquartile range (IQR) | |
| Min | Variance | |
| Max | Standard deviation | |
| | Skewness and Kurtosis | |

Being able to obtain descriptive statistics such as mean, standard deviation, skewness, kurtosis, etc. and using graphical techniques such as histograms is necessary to perform more advanced univariate analysis techniques. One such commonly used technique in intermodal freight transportation application is data fitting; that is, determining the theoretical distribution that best fit the data. A method to determine how well a theoretical distribution fits the data is to perform the goodness-of-fit (GOF) test. The GOF test statistics indicate how likely the specified theoretical distribution would produce the provided random sample.

The three most commonly used GOF tests are:

- Chi-Squared
- Kolmogorov–Smirnov (K-S)
- Anderson–Darling (A-D)

These tests essentially perform a hypothesis test of whether the sample data come from the stated distribution (null hypothesis). The null hypothesis is rejected if the computed test statistic is greater than the critical value at the desired level of confidence.

*Chi-Square Test*

The Chi-Square test statistic is computed as follows [3].

$$\chi^2 = \sum_{i=1}^{k} \frac{(O_i - E_i)^2}{E_i} \tag{11.1}$$

where $k$ is the total number of bins, $O_i$ is the observed frequency for bin $i$, and $E_i$ is the expected frequency for bin $i$ calculated by

$$E_i = n \times \big(F(x_2) - F(x_1)\big) \tag{11.2}$$

where $F$ is the cumulative distribution function (CDF) of the probability distribution being tested, $x_1$ and $x_2$ are the limits for bin $i$, and $n$ is the total number of observations. The degrees of freedom is $k$-$p$-1 where $p$ is the number of estimated parameters (including location, scale, and shape) for the sample data.

*Kolmogorov-Smirnov Test*

The Kolmogorov-Smirnov (K-S) test statistic is computed as follows [3].

$$D_n = \sup\left[ F_n(x) - \hat{F}(x) \right] \tag{11.3}$$

where $n$ is the total number of data points, $\hat{F}(x)$ is the hypothesized distribution, $F_n(x) = \frac{N_x}{n}$, and $N_x$ is the number of $X_i$ less than $x$.

*Anderson-Darling Test*

The Anderson-Darling (A-D) test statistic is computed as follows [4].

$$A^2 = -n - \frac{1}{n} \sum_{i=1}^{n} (2i-1) \times \left[ \ln F(X_i) + \ln(1 - F(X_{n-i+1})) \right] \tag{11.4}$$

where $n$ is the sample size and $F(x)$ is the theoretical CDF.

*Comments on Chi-Squared, K-S, and A-D Tests*

As stated, the value of the Chi-Squared test statistic is dependent on how the data is binned. To this end, there are no clear guidelines for selecting the size of the bins [5]. Another disadvantage of the Chi-Squared test is that it requires a large sample size. The K-S test can be used when the sample size is small. For large-sized samples, both the Chi-Squared and K-S tests yield equivalent results. The A-D test gives more weight to the tails than the K-S test. However, the critical values for the A-D test are only applicable for a few specific distributions (normal, lognormal, exponential, Weibull, extreme value type I, and logistic distributions) [6].

**Example 11.1:** Via ITS, a drayage firm has access to the processing times of the last 50 trucks at the entry gate of an intermodal terminal. For planning purposes, the firm needs to know the theoretical distribution that best fits the data. Use the Chi-Square test to determine if the data can be described by the log-normal distribution with mean of 1.3460216 and standard deviation of 0.4155127.

| Truck No. | Processing Time (min) | Truck No. | Processing Time (min) | Truck No. | Processing Time (min) |
|---|---|---|---|---|---|
| 1 | 3.5 | 18 | 7.2 | 35 | 2.3 |
| 2 | 4.7 | 19 | 2.4 | 36 | 2.6 |
| 3 | 3.7 | 20 | 6.8 | 37 | 3.1 |
| 4 | 2.9 | 21 | 4.3 | 38 | 2.1 |
| 5 | 1.3 | 22 | 5.3 | 39 | 2.8 |
| 6 | 1.7 | 23 | 1.6 | 40 | 4.6 |
| 7 | 4.2 | 24 | 4.7 | 41 | 6.7 |
| 8 | 3.7 | 25 | 5.7 | 42 | 3.9 |
| 9 | 4.0 | 26 | 3.8 | 43 | 4.8 |
| 10 | 3.7 | 27 | 4.1 | 44 | 4.9 |
| 11 | 3.4 | 28 | 5.4 | 45 | 2.5 |
| 12 | 5.6 | 29 | 7.9 | 46 | 3.9 |
| 13 | 3.5 | 30 | 8.4 | 47 | 5.3 |
| 14 | 3.2 | 31 | 3.5 | 48 | 2.5 |
| 15 | 2.7 | 32 | 4.3 | 49 | 3.2 |
| 16 | 5.9 | 33 | 5.6 | 50 | 2.1 |
| 17 | 6.1 | 34 | 6.3 | | |

**Solution:** To apply the Chi-Square test, it is necessary to put the data into bins first. For the purpose of this example, 9 bins are chosen, and the distribution of truck processing times are shown in the following figure 11.1.

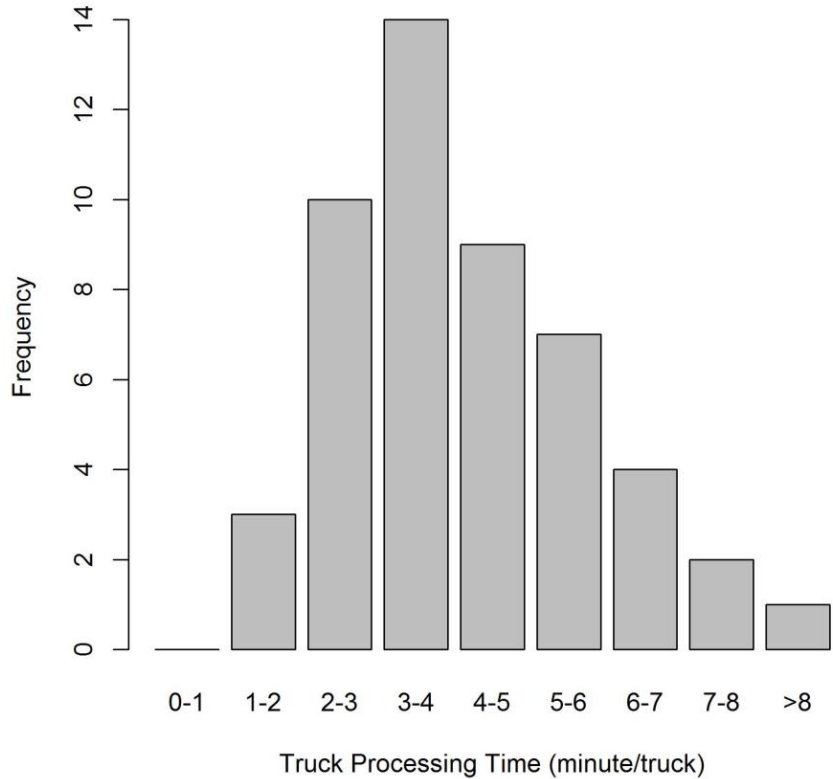

**Figure 11.1 Histogram of truck processing time**

The calculated theoretical frequency ($E_i$) and the Chi-Squared test statistics are as follows.

| Truck Processing Time | Observed Frequency ($O_i$) | Theoretical Frequency ($E_i$) | $(O_i - E_i)^2$ | $(O_i - E_i)^2 / E_i$ |
|---|---|---|---|---|
| 0.01–1 | 0 | 0.0299429 | 0.00090 | 0.02994 |
| 1–2 | 3 | 2.8731703 | 0.01609 | 0.00560 |
| 2–3 | 10 | 10.8857644 | 0.78458 | 0.07207 |
| 3–4 | 14 | 13.1414340 | 0.73714 | 0.05609 |
| 4–5 | 9 | 9.9169312 | 0.84076 | 0.08478 |
| 5–6 | 7 | 6.0680828 | 0.86847 | 0.14312 |
| 6–7 | 4 | 3.3643072 | 0.40411 | 0.12012 |
| 7–8 | 2 | 1.7816774 | 0.04766 | 0.02675 |
| > 8 | 1 | 1.9386896 | 0.88114 | 0.45450 |
| | **Σ = 50** | **Σ = 50** | | **Σ = 0.99298** |

To compute the theoretical frequency $E_i$, use Eq. 11.2.  For example, for bin 1 ($i = 1$),

$$E_1 = 50 \times \big( F(1) - F(0.01) \big)$$

The CDF of the log-normal distribution is $\Phi\left(\dfrac{\ln x - \mu}{\sigma}\right)$, where $\Phi$ is the cumulative distribution function of the *standard* normal distribution.  Thus,

$$E_1 = 50 \times \left( \Phi\left( \frac{\ln 1 - 1.34602}{0.41551} \right) - \Phi\left( \frac{\ln 0.01 - 1.34602}{0.41551} \right) \right)$$

$$E_1 = 50 \times \left( \Phi(-3.23942349) - \Phi(-14.3226160\overline{3}) \right)$$

$$E_1 = 50 \times (0.000598858 - 0) = 0.0299429$$

The values of the standard normal CDF can be obtained from tables available in any statistics textbook or by using a statistical software such as R.  To obtain the CDF value in this example, the R command `pnorm(-3.23942349)` is used.  As shown, the computed Chi-Square test statistic is 0.99298 and the critical Chi-Square value with 6 degrees of freedom ($9 - 2 - 1$) and significance level of 5% is 12.59.  Since $0.99298 < 12.59$, the null hypothesis (data come from log-normal distribution) cannot be rejected.

Notice that if there were perfect agreement between the observed and the expected frequencies, then the Chi-Square test statistic would equal zero.  The Chi-Square test statistic becomes larger as the discrepancies between the observed and expected frequencies increase.  Consequently, the null hypothesis will be rejected only if the Chi-Square test statistic is larger than critical Chi-Square value (12.59 or larger in this example).

---

The Chi-Square test can be easily performed in R.  The following commands can be used to solve Example 12.1.

```
obs_freq <- c(0, 3, 10, 14, 9, 7, 4, 2, 1)
exp_freq  <-  c(0.0299429,  2.8731703,  10.8857644,  13.1414340,  9.9169312,
6.0680828, 3.3643072, 1.7816774, 1.9386896)
chisq.test(x=obs_freq, p=exp_freq/sum(exp_freq))

Chi-squared test for given probabilities
data:  obs_freq
X-squared = 0.99298, df = 8, p-value = 0.9983
```

As highlighted, the Chi-Square test statistic is in agreement with our calculated value.

The process of determining the best-fit distribution can be done easily using R.  The following R commands can be used to check the fit of the data against a log-normal distribution.  Note that the package "fitdistrplus" is required.

```
library(fitdistrplus)
fitlogn <- fitdist(truck_processing_time,"lnorm")
gofstat(fitlogn)
```

```
Goodness-of-fit statistics
                               1-mle-lnorm
Kolmogorov-Smirnov statistic   0.07120655
Cramer-von Mises statistic     0.03071099
Anderson-Darling statistic     0.21197206
```

The parameters of the log-normal distribution can be obtained using the following R command.

<span style="color:red">summary(fitlogn)</span>

```
Fitting of the distribution ' lnorm ' by maximum likelihood
Parameters :
         estimate Std. Error
meanlog 1.3460216 0.05876237
sdlog   0.4155127 0.04155018
Loglikelihood:  -94.3359   AIC:  192.6718   BIC:  196.4958
Correlation matrix:
            meanlog          sdlog
meanlog  1.00000e+00 -8.67426e-12
sdlog   -8.67426e-12  1.00000e+00
```

Identifying the best-fit distribution and associated parameters is necessary in order to model the workflow process accurately. It should always be done instead of assuming a convenient distribution such as negative exponential.

**Bivariate Analysis**

The relationship, or lack thereof, between two variables can be determined using methods such as correlation, cross tabulation, analysis of variance, or regression. A commonly used technique in intermodal freight application is cross tabulation. Cross tabulation involves constructing a contingency table that shows the frequency of each of the variables and then using the Chi-Square test to determine whether the variables are statistically independent or if they are associated.

**Example 12.2:** For planning purposes, a customer wants to analyze the performance of its subcontractors. The table below provides data on whether carrier A or B made the delivery on time (Y = Yes, N = No). Determine if the on-time delivery of goods is associated with carrier.

| Carrier | On-Time |
|---------|---------|
| A | Y |
| B | N |
| A | N |
| A | N |
| B | Y |
| A | N |
| B | N |
| B | Y |
| A | Y |
| B | N |

| B | N |
| --- | --- |

**Solution:** Assessing the given data in the form shown above is difficult, even though it includes only 11 data pairs; a real data set is likely to have many more. Cross tabulation can be used to present the data in a more concise manner. Using carrier to represent the rows and on-time to represent the columns, the cross tabulation results are as follows.

| | | On-Time | | |
| --- | --- | --- | --- | --- |
| | | Y | N | Total |
| **Carrier** | A | 2 | 3 | 5 |
| | B | 2 | 4 | 6 |
| | Total | 4 | 7 | 11 |

Note that the cross tabulation method reduces a very large data set to just a few rows and columns (the exact number depends on the number of possible values that each variable can take).

To determine whether there is an association between on-time delivery and carrier, the Chi-Square test can be used with the following hypotheses.

$H_0$: There is no association between on-time delivery and carrier
$H_a$: There is an association between on-time delivery and carrier

Recall the Chi-Square equation (Eq. 11.1), to compute the Chi-Square test statistic we will need to compute the "expected" values. The general formula for each cell's expected frequency is:

$$E_{ij} = \frac{T_i \times T_j}{N} \tag{11.5}$$

where:
$E_{ij}$ is the expected frequency for the cell in the $i$th row and the $j$th column
$T_i$ is the total number of counts in the $i$th row
$T_j$ is the total number of counts in the $j$th column
$N$ is the total number of counts in the table

The expected frequencies are computed as follows.

| | | On-Time | | Total |
| --- | --- | --- | --- | --- |
| | | Y | N | |
| **Carrier** | A | $\frac{5 \times 4}{11} = 1.8181$ | $\frac{5 \times 7}{11} = 3.1818$ | 5 |
| | B | $\frac{6 \times 4}{11} = 2.1818$ | $\frac{6 \times 7}{11} = 3.8181$ | 6 |
| | Total | 4 | 7 | 11 |

The Chi-Square values for each cell are computed as follows.

| | | On-Time | | |
| --- | --- | --- | --- | --- |
| | | Y | N | Total |
| Carrier | A | $\dfrac{(1.8181-2)^2}{1.8181}=0.01818$ | $\dfrac{(3.1818-2)^2}{3.1818}=0.010389$ | 5 |
| | B | $\dfrac{(2.1818-2)^2}{2.1818}=0.01515$ | $\dfrac{(3.8181-2)^2}{3.8181}=0.008658$ | 6 |
| | Total | 4 | 7 | 11 |

The Chi-Square test statistic is the sum of each cell's value which is 0.05238.

The degrees of freedom are equal to $(r\text{-}1) \times (c\text{-}1)$, where $r$ is the number of rows and $c$ is the number of columns. Thus, the degrees of freedom for this problem is $(2\text{-}1) \times (2\text{-}1) = 1$. At 5% level of significance, the critical Chi-Square value is 3.84. Given that the test statistic (0.0513) is not greater than or equal to critical value (3.84), we cannot reject the null hypothesis. That is, there is no association between on-time delivery and carrier.

---

The cross tabulation analysis can be easily performed in R with the "rpivotTable" package. The following commands would give the cross tabulation table.

```
library(rpivotTable)
rpivotTable(Data,rows="Carrier", cols=c("On-Time"))
```

The Chi-Square test results can be obtained using the following R commands.

```
result <- xtabs(~Carrier + On-Time, Data)
summary(result)

Call: xtabs(formula = ~Carrier + On-Time, data = Data)
Number of cases in table: 11
Number of factors: 2
Test for independence of all factors:
        Chisq = 0.05238, df = 1, p-value = 0.819
        Chi-squared approximation may be incorrect
```

As highlighted, the Chi-Square test statistic and degrees of freedom are in agreement with our calculated values. Note that due to the small sample size, R issued a warning that the test results may not be valid.

## PREDICTIVE DATA ANALYTICS

### Bivariate Analysis

The previous section illustrated the cross tabulation technique to determine if there is an association or relationship between two variables. If a relationship is established, a common practice is to use this relationship for prediction. It is typically assume that the relationship is linear. This assumption provides a convenient and tractable set of equations known as simple linear regression.

$$y = b_0 + b_1 x \tag{11.6}$$

where $y$ is the dependent variable, $x$ is the independent (or explanatory) variable, $b_0$, is the y-intercept, and $b_1$ is the slope of the regression line.

$b_0$ and $b_1$ can be computed as follows

$$b_0 = \overline{y} - b_1 \overline{x} \tag{11.7}$$

$$b_1 = \frac{SS(xy)}{SS(x)} \tag{11.8}$$

The notation $SS(x)$ and $SS(xy)$ are defined as

$$SS(x) = \sum x^2 - \frac{\left(\sum x\right)^2}{n} \tag{11.9}$$

$$SS(xy) = \sum xy - \frac{\left(\sum x\right)\left(\sum y\right)}{n} \tag{11.10}$$

**Example 11.3:** From an integrated freight database enabled by ITS, a metropolitan planning organization has access to the following data which contain information about the fleet size and trips made by different drayage firms operating near a seaport. Use the survey data to establish the linear relationship between the number of trucks available and the number of port trips made in a day and use this relationship to predict how many port trips will be made by a new drayage firm that is expected to have 10 trucks.

| Drayage Firm | No. of Port Trips Made in a Day | No. of Trucks Available |
|:---:|:---:|:---:|
| 1 | 3 | 4 |
| 2 | 1 | 2 |
| 3 | 2 | 3 |
| 4 | 4 | 4 |
| 5 | 3 | 2 |
| 6 | 2 | 4 |
| 7 | 6 | 8 |
| 8 | 4 | 6 |
| 9 | 5 | 6 |
| 10 | 2 | 2 |

**Solution:** Compute the $SS(x)$ quantities as follows.

| No. of Port Trips Made in a Day ($y$) | No. of Trucks Available ($x$) | $x^2$ | $xy$ |
|:---:|:---:|:---:|:---:|
| 3 | 4 | 16 | 12 |
| 1 | 2 | 4 | 2 |
| 2 | 3 | 9 | 6 |
| 4 | 4 | 16 | 16 |
| 3 | 2 | 4 | 6 |
| 2 | 4 | 16 | 8 |
| 6 | 8 | 64 | 48 |
| 4 | 6 | 36 | 24 |
| 5 | 6 | 36 | 30 |
| 2 | 2 | 4 | 4 |
| $\Sigma = 32$ | $\Sigma = 41$ | $\Sigma = 205$ | $\Sigma = 156$ |

$$SS(x) = \sum x^2 - \frac{\left(\sum x\right)^2}{n} = 205 - \frac{(41)^2}{10} = 36.9$$

$$SS(xy) = \sum xy - \frac{\left(\sum x\right)\left(\sum y\right)}{n} = 156 - \frac{(41)(32)}{10} = 24.8$$

$$b_1 = \frac{SS(xy)}{SS(x)} = \frac{24.8}{36.9} = 0.6721$$

$$b_0 = \overline{y} - b_1\overline{x} = \frac{32}{10} - (0.6721)\left(\frac{41}{10}\right) = 0.4444$$

Thus, the equation that depicts the linear relationship between the number of trucks available and number of port trips made in a day is

$$y = 0.4444 + 0.6721x$$

To predict the number of port trips to be made by the drayage firm, use the above equation with $x$ = 10. That is,

$$y = 0.4444 + 0.6721(10) = 7.1653$$

---

The following R commands can be used to obtain the simple linear regression model for the problem in Example 11.3.

```
port_trips <- c(3, 1, 2, 4, 3, 2, 6, 4, 5, 2)
trucks_avail <-c(4, 2, 3, 4, 2, 4, 8, 6, 6, 2)
```

```
linear_model <- lm(port_trips ~ trucks_avail)
summary(linear_model)

Call:
lm(formula = port_trips ~ truck_avail)

Residuals:
     Min       1Q   Median       3Q      Max
-1.13279 -0.47290  0.02304  0.44512  1.21138

Coefficients:
            Estimate Std. Error t value Pr(>|t|)
(Intercept)   0.4444     0.5852   0.759 0.469391
truck_avail   0.6721     0.1293   5.199 0.000823 ***
---
Signif. codes:  0 '***' 0.001 '**' 0.01 '*' 0.05 '.' 0.1 ' ' 1

Residual standard error: 0.7852 on 8 degrees of freedom
Multiple R-squared:  0.7717,    Adjusted R-squared:  0.7431
F-statistic: 27.03 on 1 and 8 DF,  p-value: 0.0008229
```

As highlighted, the R generated values for $b_0$ and $b_1$ are in agreement with our calculated values.

To use the simple regression model to predict the number of port trips if the number of trucks available is 10, the following R commands can be used.

```
newdata = data.frame(trucks_avail=10)
predict(linearmodel, newdata)
```

The result is 7.1653 which is the same as our calculated value.

How well the straight line fits the data can be determined from the R-squared value, known as the coefficient of determination. The R-squared value ranges from 0 to 1, where 1 indicates a perfect model. The coefficient of determination is defined as follows.

$$R^2 = \frac{SS(y) - SSE}{SS(y)} = 1 - \frac{SSE}{SS(y)} \tag{11.11}$$

Where SSE is the sum of the squared deviations of the points to the least squares line and $SS(y)$ is defined as

$$SS(y) = \sum y^2 - \frac{\left(\sum y\right)^2}{n} \tag{11.12}$$

The calculation of the coefficient of determination of the simple regression model estimated for Example 11.3 is left as an exercise for the reader.

Generally, to measure the strength of the linear relationship, the Pearson's product-moment correlation coefficient can be used. It is defined as follows.

$$r = \frac{SS(xy)}{\sqrt{SS(x)SS(y)}}$$ (11.13)

*Properties of the Pearson's Correlation Coefficient, r:*

1. The value of $r$ is always between $-1$ and $+1$ inclusive.
2. $r$ has the same sign as $b_1$, the slope of the least square line.
3. $r$ is near $+1$ when the data points fall close to the straight line with positive slope.
4. $r$ is near $-1$ when the data points fall close to the straight line with negative slope.
5. If all the data points fall exactly on the straight line with positive slope, then $r = +1$.
6. If all the data points fall exactly on the straight line with negative slope, then $r = -1$.
7. A value of $r$ near 0 indicates little or no linear relationship between $y$ and $x$.

**Example 11.4:** Verify that the value of the Pearson's correlation coefficient squared is the same value as the coefficient of determination in Example 11.3.

**Solution:** $SS(x)$ and $SS(xy)$ have been computed previously in Example 11.3.

$$SS(xy) = \sum xy - \frac{\left(\sum x\right)\left(\sum y\right)}{n} = 156 - \frac{(41)(32)}{10} = 24.8$$

$$SS(x) = \sum x^2 - \frac{\left(\sum x\right)^2}{n} = 205 - \frac{(41)^2}{10} = 36.9$$

$SS(y)$ is computed as follows.

$$SS(y) = \sum y^2 - \frac{\left(\sum y\right)^2}{n} = 124 - \frac{(32)^2}{10} = 21.6$$

Applying Eq. 11.14, we obtain

$$r = \frac{SS(xy)}{\sqrt{SS(x)SS(y)}} = \frac{24.8}{\sqrt{(36.9)(21.6)}} = 0.8784$$

and thus, $r^2 = 0.7717$, which is equal to the coefficient of determination.

**Multivariate Analysis**

In intermodal freight transportation applications, oftentimes the variable of interest (dependent variable) is associated or related to two or more explanatory variables. In this case, instead of a simple regression model, we have a multiple regression model. A multiple linear regression model with two explanatory variables has the following form.

$$y = b_0 + b_1 x_1 + b_2 x_2 \tag{11.14}$$

where $b_0$ is the y-intercept, $b_1$ is the change in $y$ for each 1 unit change in $x_1$, and $b_2$ is the change in $y$ for each 1 unit change in $x_2$.

$b_0$, $b_1$, and $b_2$ can be computed as follows.

$$b_1 = \left( \frac{r_{y,x_1} - r_{y,x_2} \, r_{x_1,x_2}}{1 - \left(r_{x_1,x_2}\right)^2} \right) \left( \frac{SD_y}{SD_{x_1}} \right) \tag{11.15}$$

$$b_2 = \left( \frac{r_{y,x_2} - r_{y,x_1} \, r_{x_1,x_2}}{1 - \left(r_{x_1,x_2}\right)^2} \right) \left( \frac{SD_y}{SD_{x_2}} \right) \tag{11.16}$$

$$b_0 = \overline{y} - b_1 \overline{x_1} - b_2 \overline{x_2} \tag{11.17}$$

where $SD_y$ = standard deviation of $y$, $SD_{x_1}$ = standard deviation of $x_1$, $SD_{x_2}$ = standard deviation of $x_2$, $r_{y,x_1}$ = correlation between $y$ and $x_1$, $r_{y,x_2}$ = correlation between $y$ and $x_2$, and $r_{x_1,x_2}$ = correlation between $x_1$ and $x_2$.

The correlation values are defined as follows.

$$r_{y,x_1} = \frac{n \sum y \times x_1 - \sum y \sum x_1}{\sqrt{n \sum x_1^2 - (\sum x_1)^2} \, \sqrt{n \sum y^2 - (\sum y)^2}} \tag{11.18}$$

$$r_{y,x_2} = \frac{n \sum y \times x_2 - \sum y \sum x_2}{\sqrt{n \sum x_2^2 - (\sum x_2)^2} \, \sqrt{n \sum y^2 - (\sum y)^2}} \tag{11.19}$$

$$r_{x_1,x_2} = \frac{n \sum x_1 \times x_2 - \sum x_1 \sum x_2}{\sqrt{n \sum x_1^2 - (\sum x_1)^2} \, \sqrt{n \sum x_2^2 - (\sum x_2)^2}} \tag{11.20}$$

**Example 11.5:** Using the terminal webcams and image processing techniques, a drayage firm was able to obtain the following data at the entry gate of a marine container terminal. Use this data to develop a multiple linear regression model with truck queueing time as the dependent variable, and gate processing time and queue length as explanatory variables. Then use the developed model to predict the truck queueing time when gate processing time is 5 minutes and there are 6 trucks in the queue.

| Queue Length (No. of Trucks) | Gate Processing Time (min) | Truck Queueing Time (min) |
|:---:|:---:|:---:|
| 1 | 2 | 2 |
| 3 | 2 | 5 |
| 2 | 3 | 7 |
| 4 | 8 | 15 |
| 2 | 4 | 10 |

**Solution:** Compute the mean and standard deviation of the variables as follows.

| Variables | Mean | Standard Deviation, SD |
|:---|:---:|:---:|
| Truck queueing time ( $y$ ) | 7.8 | 4.9699 |
| Queue length ( $x_1$ ) | 2.4 | 1.1402 |
| Gate processing time ( $x_2$ ) | 3.8 | 2.4900 |

Next, compute the correlation values as follows.

| $y$ | $x_1$ | $x_2$ | $y^2$ | $x_1^2$ | $x_2^2$ | $y*x_1$ | $y*x_2$ | $x_1*x_2$ |
|:---:|:---:|:---:|:---:|:---:|:---:|:---:|:---:|:---:|
| 2 | 1 | 2 | 4 | 1 | 4 | 2 | 4 | 2 |
| 5 | 3 | 2 | 25 | 9 | 4 | 15 | 10 | 6 |
| 7 | 2 | 3 | 49 | 4 | 9 | 14 | 21 | 6 |
| 15 | 4 | 8 | 225 | 16 | 64 | 60 | 120 | 32 |
| 10 | 2 | 4 | 100 | 4 | 16 | 20 | 40 | 8 |
| $\Sigma = 39$ | $\Sigma = 12$ | $\Sigma = 19$ | $\Sigma = 403$ | $\Sigma = 34$ | $\Sigma = 97$ | $\Sigma = 111$ | $\Sigma = 195$ | $\Sigma = 54$ |

Applying Eq. 11.18 (to obtain correlation between truck queueing time and queue length), we have

$$r_{y,x_1} = \frac{n\sum y \times x_1 - \sum y \sum x_1}{\sqrt{n\sum x_1^2 - (\sum x_1)^2}\sqrt{n\sum y^2 - (\sum y)^2}} = \frac{5 \times 111 - 39 \times 12}{\sqrt{5*34 - (12)^2}\sqrt{5 \times 403 - (39)^2}} = 0.7677$$

Applying Eq. 11.19 (to obtain correlation between truck queueing time and gate processing time), we have

$$r_{y,x_2} = \frac{n\sum y \times x_2 - \sum y \sum x_2}{\sqrt{n\sum x_2^2 - (\sum x_2)^2}\sqrt{n\sum y^2 - (\sum y)^2}} = 0.9455$$

Applying Eq. 11.20 (to obtain correlation between queue length and gate processing time), we have

$$r_{x_1, x_2} = \frac{n \sum x_1 \times x_2 - \sum x_1 \sum x_2}{\sqrt{n \sum x_1^2 - (\sum x_1)^2} \sqrt{n \sum x_2^2 - (\sum x_2)^2}} = 0.7397$$

The regression coefficients are calculated as follows.

$$b_1 = \left( \frac{r_{y,x_1} - r_{y,x_2} \, r_{x_1,x_2}}{1 - (r_{x_1,x_2})^2} \right) \left( \frac{SD_y}{SD_{x_1}} \right) = \frac{0.7677 - 0.9455 \times 0.7397}{1 - (0.7397)^2} \times \frac{4.9699}{1.1402} = 0.6575$$

$$b_2 = \left( \frac{r_{y,x_2} - r_{y,x_1} \, r_{x_1,x_2}}{1 - (r_{x_1,x_2})^2} \right) \left( \frac{SD_y}{SD_{x_2}} \right) = \frac{0.9455 - 0.7677 \times 0.7397}{1 - (0.7397)^2} \times \frac{4.9699}{2.4900} = 1.6644$$

$$b_0 = \bar{y} - b_1 \bar{x_1} - b_2 \bar{x_2} = 7.8 - 0.6575 \times 2.4 - 1.6644 \times 3.8 = -0.1027$$

Lastly, applying Eq. 11.14, we have

$$y = -0.1027 + 0.6575 x_1 + 1.6644 x_2$$

To predict the truck queueing time, use the above equation with $x_1 = 6$ and $x_2 = 5$. That is,

$$y = -0.1027 + 0.6575(6) + 1.6644(5) = 12.1643 \text{ minutes}$$

---

The following R commands can be used to obtain the multiple linear regression model for the problem presented in Example 11.5.

```
queue_length <- c(1,3,2,4,2)
gate_time <- c(2,2,3,8,4)
queueing_time <- c(2,5,7,15,10)
mreg <- lm(queueing_time ~ queue_length + gate_time)
summary(mreg)

Call:
lm(formula = queueing_time ~ queue_length + gate_time)

Residuals:
       1        2        3        4        5
-1.8836  -0.1986   0.7945  -0.8425   2.1301

Coefficients:
             Estimate Std. Error t value Pr(>|t|)
(Intercept)   -0.1027     2.4883  -0.041    0.971
queue_length   0.6575     1.4177   0.464    0.688
gate_time      1.6644     0.6492   2.564    0.124

Residual standard error: 2.176 on 2 degrees of freedom
Multiple R-squared:  0.9042,  Adjusted R-squared:  0.8084
F-statistic: 9.438 on 2 and 2 DF,  p-value: 0.09581
```

As highlighted, the R generated values for $b_0$, $b_1$, and $b_2$ are in agreement with our calculated values.

To use the multiple regression model to predict the truck queuing time when the gate processing time is 5 minutes and there are 6 trucks in the queue, the following R commands can be used.

```
newdata = data.frame(gate_time=5, queue_length=6)
predict(mreg, newdata)
```

The result is 12.1643 which is the same as our calculated value.

For problems with more than two explanatory variables, determining the coefficients of the multiple linear regression model using the analytical equations could be tedious. In practice, statistical software is the method of choice to obtain the model coefficients and the coefficient of determination (R-Squared). In R, adding another explanatory variable is straightforward. One just need to add another plus sign and the variable name. For example, suppose in Example 11.5 we have a third explanatory variable: transaction type. The R command with three variables would be

```
mreg <- lm(queueing_time ~ queue_length + gate_time + transaction_type)
```

**Fuzzy Regression**

Although multiple regression can be applied to a variety of problems, there are situations when they are not appropriate. These situations include: 1) the data set is too small, 2) the error is not normally distributed, 3) there is vagueness in the relationship between the explanatory and dependent variables, 4) there is ambiguity associated with the event being modeled, and 5) the linearity assumption is inappropriate. Fuzzy regression can be used in these situations. It is based on the fuzzy set theory and was introduced by Tanaka et al. in 1982 [7]. In this method, the deviations between observed and predicted value reflect the vagueness of the data pattern. The pattern of the data is expressed by the model's fuzzy parameters, which can be solved by linear programming. The objective of the linear program is to minimize the fuzzy deviations subject to some degree of membership fit constraints. Let us consider a case where the dependent variable is $y$ and the explanatory variables are $x_1$ and $x_2$. To formulate the fuzzy regression model, three new variables need to be defined: $X_0$, $X_1$, and $X_2$. The relationship between $X_p$ and $x_p$ is $X_p = x_{ip}$ for $p = 0, 1, 2$ and $i = 1, \ldots, n$. The fuzzy regression model can be written as

$$\tilde{y} = \tilde{A}_0 X_0 + \tilde{A}_1 X_1 + \tilde{A}_2 X_2 \tag{11.21}$$

where, $\tilde{A}_0, \tilde{A}_1, \tilde{A}_2 =$ Fuzzy coefficients

$\qquad X_0 = x_{i0} = 1$, for $i = 1, \ldots, n$

$X_1 = x_{i1}$, for $i = 1, \ldots, n$

$X_2 = x_{i2}$, for $i = 1, \ldots, n$

$n$ = total number of observations.

Representing each fuzzy number by its fuzzy center ($a_k$) and radius ($c_k$), we have the following equation.

$$\langle y_a, y_c \rangle = \langle a_0, c_0 \rangle + \langle a_1, c_1 \rangle X_1 + \langle a_2, c_2 \rangle X_2 \tag{11.22}$$

The linear program to determine the fuzzy parameters is shown below.

$$\min \; Z = c_0 \sum_{i=1}^{n} x_{i0} + c_1 \sum_{i=1}^{n} x_{i1} + c_2 \sum_{i=1}^{n} x_{i2} \tag{11.23}$$

subject to,

$$\sum_{k=0}^{2} a_k x_{ik} + (1-h) \sum_{k=0}^{2} c_k x_{ik} \geq y_i, \quad \forall i = 1, \ldots, n \tag{11.24}$$

$$\sum_{k=0}^{2} a_k x_{ik} - (1-h) \sum_{k=0}^{2} c_k x_{ik} \leq y_i, \quad \forall i = 1, \ldots, n \tag{11.25}$$

where $c_k \geq 0$, $a_k \in R$, $x_{i0} = 1$, $0 \leq h \leq 1$, and $h$ = certain factor.

The values of $a_k$ and $c_k$ can be obtained by solving the linear program. These values can then be substituted into Eq. 11.23 for prediction purposes.

**Example 11.6:** Apply fuzzy regression technique to the data presented in Example 11.5.

**Solution:** Applying the linear program with the certain factor $h = 0.9$, we have

$$\min \; Z = c_0 \sum_{i=1}^{5} x_{i0} + c_1 \sum_{i=1}^{5} x_{i1} + c_2 \sum_{i=1}^{5} x_{i2} = 5c_0 + 12c_1 + 19c_2$$

subject to,

$a_0 + a_1 + 2a_2 + (1 - 0.9)(c_0 + c_1 + 2c_2) \geq 2$

$a_0 + a_1 + 2a_2 - (1 - 0.9)(c_0 + c_1 + 2c_2) \leq 2$

$a_0 + 3a_1 + 2a_2 + (1 - 0.9)(c_0 + 3c_1 + 2c_2) \geq 5$

$a_0 + 3a_1 + 2a_2 - (1 - 0.9)(c_0 + 3c_1 + 2c_2) \leq 5$

$a_0 + 2a_1 + 3a_2 + (1 - 0.9)(c_0 + 2c_1 + 3c_2) \geq 7$

$$a_0 + 2a_1 + 3a_2 - (1 - 0.9)(c_0 + 2c_1 + 3c_2) \leq 7$$

$$a_0 + 4a_1 + 8a_2 + (1 - 0.9)(c_0 + 4c_1 + 8c_2) \geq 15$$

$$a_0 + 4a_1 + 8a_2 - (1 - 0.9)(c_0 + 4c_1 + 8c_2) \leq 15$$

$$a_0 + 2a_1 + 4a_2 + (1 - 0.9)(c_0 + 2c_1 + 4c_2) \geq 10$$

$$a_0 + 2a_1 + 4a_2 - (1 - 0.9)(c_0 + 2c_1 + 4c_2) \leq 10$$

$$c_0, c_1, c_2 \geq 0$$

$$a_0, a_1, a_2 \in R$$

The above linear program can be solved in R with the "lpSolveAPI" package. The following R commands can be used to obtain the decision variables ($a_k$ and $c_k$).

```
library(lpSolveAPI)
lp.truck <- make.lp(0,6)
lp.control(lp.truck, sense="min")
set.objfn(lp.truck, c(0, 0, 0, 5, 12, 19))
add.constraint(lp.truck, c(1, 1, 2, 0.1, 0.1, 0.2), ">=", 2)
add.constraint(lp.truck, c(1, 1, 2, -0.1, -0.1, -0.2), "<=", 2)
add.constraint(lp.truck, c(1, 3, 2, 0.1, 0.3, 0.2), ">=", 5)
add.constraint(lp.truck, c(1, 3, 2, -0.1, -0.3, -0.2), "<=", 5)
add.constraint(lp.truck, c(1, 2, 3, 0.1, 0.2, 0.3), ">=", 7)
add.constraint(lp.truck, c(1, 2, 3, -0.1, -0.2, -0.3), "<=", 7)
add.constraint(lp.truck, c(1, 4, 8, 0.1, 0.4, 0.8), ">=", 15)
add.constraint(lp.truck, c(1, 4, 8, -0.1, -0.4, -0.8), "<=", 15)
add.constraint(lp.truck, c(1, 2, 4, 0.1, 0.2, 0.4), ">=", 10)
add.constraint(lp.truck, c(1, 2, 4, -0.1, -0.2, -0.4), "<=", 10)
set.bounds(lp.truck, lower=c(-Inf, -Inf, -Inf), upper=c(Inf, Inf, Inf),
columns=1:3)
set.bounds(lp.truck, lower=c(0, 0, 0), upper=c(Inf, Inf, Inf), columns=4:6)
solve(lp.truck)
get.objective(lp.truck)
get.variables(lp.truck)
```

At optimality, the values of the decision variables are

$a_0 = -2.333$
$a_1 = 1.5$
$a_2 = 1.875$
$c_0 = 0$
$c_1 = 0$
$c_2 = 4.583$

And thus the fuzzy regression coefficients are

$a_0 - c_0 = -2.333$
$a_1 - c_1 = 1.5$
$a_2 - c_2 = -2.708$
$a_0 + c_0 = -2.333$
$a_1 + c_1 = 1.5$
$a_2 + c_2 = 6.458$

The fuzzy regression model does not provide a specific value for queueing time, but rather a range for the queueing time. Using $x_1 = 6$ and $x_2 = 5$, the lower predicted truck queueing time is $= -2.333 + 1.5(6) - 2.708(5) = -6.873$ minutes. The upper predicted truck queuing time is $= -2.333 + 1.5(6) + 6.458(5) = 38.957$ minutes. Note that the predicted values are not bounded; thus, the minimum predicted values may be negative.

To compare the above predicted truck queueing time with that of multiple linear regression, the average predicted values of this method (which is the middle value of the predicted range) can be used. For this example, the average truck queueing time is 16.04 minutes.

## SUMMARY AND CONCLUSIONS

Transportation planners and analysts need to be equipped with the data analytic techniques to be able to understand and visualize intermodal freight transportation data. This chapter has presented the commonly used techniques: data fitting, cross tabulations, linear regression, and fuzzy regression. These techniques cover the entire spectrum of univariate, bivariate, and multivariate analyses. In addition to illustrating how to apply these techniques through relatively simple examples, this chapter also showed how they can be applied using the statistical software R. Additional exercises are provided for those who wish to apply these techniques to more complex problems.

Current communications and information technology and ITS systems are enabling third-party logistics providers, trucking companies, railroads, ocean carriers, and terminal operators to work together to provide a cost-effective and efficient delivery of freight from their origins to their destinations. The opportunity for information sharing will improve local and regional intermodal operations. However, it will also pose a challenge to stakeholders due to the data sources being large and exist in different formats, resolution, and time scales. To be successful in future ITS-enabled intermodal freight transportation system, these stakeholders will need to apply the data analytics techniques discussed in this chapter.

## EXERCISE PROBLEMS

**11.1** Apply the Anderson-Darling (A-D) test for the data given in Example 11.1.

**11.2** The following table shows the results of a survey conducted by a terminal operator to determine if truck queuing time at the entry gate is associated with weather. Truck queuing time is classified as L (less than 30 minutes) or H (30 minutes or longer), and weather is classified as N (no rain), L (light rain), M (moderate rain), and H (heavy rain). Determine if truck queuing time is associated with weather.

| Truck No. | Delay | Rain | Truck No. | Delay | Rain | Truck No. | Delay | Rain |
|-----------|-------|------|-----------|-------|------|-----------|-------|------|
| 1 | L | L | 18 | L | L | 35 | L | L |
| 2 | L | N | 19 | H | N | 36 | H | L |
| 3 | L | N | 20 | H | H | 37 | L | M |
| 4 | L | N | 21 | H | N | 38 | L | H |
| 5 | H | M | 22 | H | M | 39 | L | H |
| 6 | H | H | 23 | H | L | 40 | H | L |
| 7 | L | M | 24 | H | H | 41 | H | L |
| 8 | L | L | 25 | H | L | 42 | L | N |
| 9 | L | M | 26 | L | M | 43 | H | H |
| 10 | H | H | 27 | L | L | 44 | H | M |
| 11 | H | H | 28 | H | H | 45 | L | N |
| 12 | L | N | 29 | H | N | 46 | L | M |
| 13 | L | H | 30 | H | M | 47 | H | M |
| 14 | H | L | 31 | H | M | 48 | L | H |
| 15 | L | N | 32 | L | L | 49 | H | N |
| 16 | H | H | 33 | H | M | 50 | H | H |
| 17 | L | L | 34 | L | L | | | |

**11.3** Calculate the coefficient of determination ($R^2$) for the simple regression model estimated in Example 11.3.

**11.4** A class I rail company wants to determine the relationship between the total transportation costs of an intermodal container ($1,000) and the explanatory variables: distance between shipment origin and destination (miles), number of intermodal transfer and shipment delivery time (days). The data for 10 different types of commodities are shown in the table below.

| Commodity | Total Transportation Costs ($1,000/container) | Distance between Origin and Destination (miles) | Number of Intermodal Transfer | Delivery Time (days) |
|-----------|-----------------------------------------------|-------------------------------------------------|-------------------------------|----------------------|
| 1 | 5.0 | 500 | 2 | 7 |
| 2 | 4.8 | 550 | 2 | 7 |
| 3 | 6.5 | 600 | 3 | 14 |
| 4 | 6.5 | 650 | 3 | 7 |
| 5 | 5.3 | 550 | 2 | 14 |
| 6 | 6.5 | 700 | 4 | 7 |
| 7 | 7.0 | 800 | 4 | 7 |
| 8 | 6.8 | 600 | 2 | 14 |
| 9 | 7.5 | 700 | 3 | 7 |

| 10 | 5.7 | 550 | 2 | 7 |
|---|---|---|---|---|

(i) Write the multiple regression model, (ii) Estimate the value of the coefficient of determination ($R^2$), and (iii) Use the developed model to predict the transportation costs of an intermodal container that is transported 700 miles in 14 days with 3 intermodal transfers.

**11.5** Using the data presented in Problem 11.4, develop a fuzzy regression model where the dependent variable is total transportation cost and the explanatory variables are distance between shipment origin and destination and the number of intermodal transfers.

## SOLUTION TO EXERCISE PROBLEMS

**11.1** $A^2$ = 0.21197206, Critical value = 0.740230338

**11.2** $\chi^2$ = 3.28556, Critical value = 7.81

**11.3** $R^2$ = 0.7717

**11.4** (i) Costs = –0.4836 + 0.0110*Distance – 0.2957*Transfer + 0.0684*Delivery time; (ii) 0.73; and (iii) $7,286.21/container.

**11.5** $a_0$ = 0.36, $a_1$ = 0.0088, $a_2$ = 0.19, $c_0$ = 7.8, $c_1$ = 0, and $c_2$ = 0.